\documentclass[aps,prl,twocolumn,superscriptaddress,showpacs]{revtex4-1} 
\usepackage{amsmath}
\usepackage{graphicx,}
\usepackage{dcolumn}
\usepackage{amsmath}
\usepackage{amssymb}
\usepackage{float}
\usepackage{epstopdf}
\usepackage{wasysym}
\usepackage{color}
\setcitestyle{super}
\begin{document}	
	\author{Ahmad K. Omar}
	\affiliation{Division of Chemistry and Chemical Engineering, California Institute of Technology, Pasadena, California 91125, USA}
	\author{Yanze Wu}
	\affiliation{Division of Chemistry and Chemical Engineering, California Institute of Technology, Pasadena, California 91125, USA}
	\affiliation{School of Chemistry and Chemical Engineering, Shanghai Jiao Tong University, Shanghai 200240, China}
	\author{Zhen-Gang Wang}
	\email[]{zgw@caltech.edu}
	\author{John F. Brady}
	\email[]{jfbrady@caltech.edu}
	\affiliation{Division of Chemistry and Chemical Engineering, California Institute of Technology, Pasadena, California 91125, USA}
	
	\title{Swimming to Stability: Structural and Dynamical Control \textit{via} Active Doping}
	
\begin{abstract}
External fields can decidedly alter the free energy landscape of soft materials and can be exploited as a powerful tool for the assembly of targeted nanostructures and colloidal materials. Here, we use computer simulations to demonstrate that nonequilibrium \textit{internal fields} or forces -- forces that are generated by driven components within a system -- in the form of active particles can precisely modulate the dynamical free energy landscape of a model soft material, a colloidal gel. Embedding a small fraction of active particles within a gel can provide a unique pathway for the dynamically frustrated network to circumvent the kinetic barriers associated with reaching a lower free energy state through thermal fluctuations alone. Moreover, by carefully tuning the active particle properties (the propulsive swim force and persistence length) in comparison to those of the gel, the active particles may induce depletion-like forces between the constituent particles of the gel despite there being \textit{no geometric size asymmetry} between the particles. These resulting forces can rapidly push the system toward disparate regions of phase space. Intriguingly, the state of the material can be altered by tuning macroscopic transport properties such as the solvent viscosity. Our findings highlight the potential wide-ranging structural and kinetic control facilitated by varying the dynamical properties of a remarkably small fraction of driven particles embedded in a host material.
\end{abstract}

\maketitle

Advances in materials chemistry have opened new and exciting vistas for designing colloidal and nanoparticle building blocks, resulting in materials endowed with unique structures and accompanying properties.~\cite{Glotzer2007, Grzelczak2010, Sacanna2011, Boles2016} Particle shape~\cite{Damasceno2012, Anders2014} and surface chemistry (\textit{e.g.},~decorating the particle surface with sticky ``patches"~\cite{Zhang2004, Jiang} of organic or biological molecules~\cite{Mirkin1996, Wang2012, Obana2017}) allow for exquisite control of the directionality of particle interactions, unlocking a multitude of structures with wide-ranging symmetries. The attractive particle interactions that are often used to stabilize these structures may also introduce a free energy landscape replete with barriers associated with particle rearrangement which can significantly delay or even preclude reaching the globally stable configuration. Indeed, the traditional pathways for homogeneous suspensions of particles to phase separate into dilute and dense phases, including spinodal decomposition and nucleation, can be exceedingly long and in practice can be the limiting factor in the experimental observation of ordered colloidal and nanoparticle-based materials.~\cite{Fan, Anderson2002}

The classical system that demonstrates this issue of kinetic frustration is perhaps also the simplest: a colloidal suspension of spheres with short-range isotropic attractive interactions. While the equilibrium phase diagram of this system predicts fluid-solid coexistence for interaction strengths as little as a few times the thermal energy $k_BT$, in practice the system is found to form a long-lived space-spanning network, a gel, for nearly the entire coexistence region.~\cite{Lu2008} This gel state is argued to be a result of the glassy dynamics of the solid phase which hinders the diffusion and coalescence of the bicontinuous dense strands formed during spinodal decomposition.~\cite{Lu2008} Even upon reducing the isotropic attraction to discrete sticky sites, it can take years for a colloidal system to complete the phase separation process.~\cite{Ruzicka2010}

Nonequilibrium protocols provide an opportunity to overcome the naturally occurring kinetic barriers.  A classic approach is simple thermal annealing; temporally controlling the presence of potential energy barriers by modulating the system temperature can allow for particles to search for stable configurations before the barriers have fully set in. Recently, Swan and co-workers explored temporally varying the interaction potential (\textit{e.g.}~for systems in which the interaction energy can be modulated with an external field) and found improved colloidal crystallization rates by periodically ``toggling" the interactions off and on.~\cite{Swan2012, Swan2014, Risbud2015, Sherman2016, Sherman2018} Applied shear or stress has also been examined as a means by which to reduce kinetic barriers by allowing dense or yield-stress materials to ``fluidize" or yield.~\cite{Liu1998, Trappe2001} Indeed, simulations of colloidal gels in a variety of deformation protocols have revealed an increase in the rate of gel coarsening or phase separated-like states at certain applied rates/stresses.~\cite{Colombo2014, Koumakis2015, Landrum2016, Johnson2018} In practice, however, while deformation-induced structures may resemble the target-structure, deformation protocols naturally break rotational symmetry which could be an essential feature of the target structure. Indeed, the development of protocols that truly preserve the underlying (equilibrium) globally stable configuration while providing a viable kinetic pathway to this configuration is a difficult balance and remains an outstanding challenge in colloidal and materials science.

The above examples involve the application of \textit{external} fields to circumvent kinetic barriers. Broadly, these protocols act to inject the necessary free energy (through the system boundary) to either traverse the existing barriers or create entirely new pathways for phase separation. However, these free-energy injections need not come from an external source. The recent focus on so-called active particles~\cite{Bechinger2016, Takatori2016, Mallory2018} -- particles that self-propel or ``swim" through the conversion of chemical energy -- motivates the idea of doping a material with a small fraction of such particles which can act as \textit{internal} sources of free energy that can locally drive a material over kinetic hurdles. A complementary mechanical viewpoint is that active particles - just as shear and other deformation protocols -- can act as a source of stress by virtue of their unique ``swim pressure"~\cite{Takatori2014, Fily2014, Solon2015a} which, on the colloidal scale, can be the dominant source of stress. 

The use of active particles in material design has previously been explored; however most of these studies have focused on suspensions in which \textit{all} particles are active.~\cite{Redner2013a, Prymidis2015, Mallory2016, Mallory2017} To date, a few studies have sought to characterize the influence of a small fraction of active particles on the dynamical and structural landscape of the host material.~\cite{Massana-Cid2018} These studies have revealed intriguing results: simulations~\cite{Ni2014} have shown that embedding a hard-sphere glass with active particles can improve crystallization rates while experiments on slightly more dilute hard-sphere systems demonstrated that active particles improve the coalescence of large passive grains. Other studies that have explored mixtures of active and ``passive" particles~\cite{Stenhammar2015, Takatori2015a} have largely focused on regimes in which passive particles perturb the underlying properties/behavior of the active particles (such as motility-induced phase separation or MIPS~\cite{Cates2015}), the inverse problem to the active doping idea explored here.

\begin{figure}
	\centering
	\includegraphics[width=0.48\textwidth,keepaspectratio,clip]{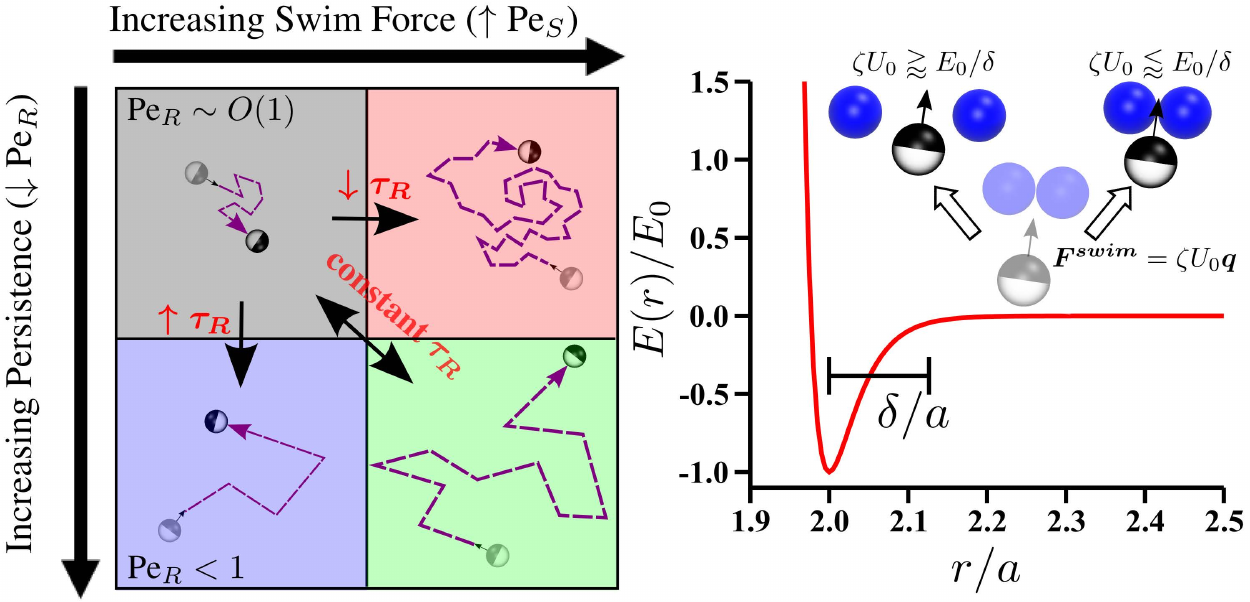}
	\caption{{\protect\small{Physical schematic (left) of the key parameters characterizing active particle motion: persistence and swim force quantified by $\text{Pe}_R$ and $\text{Pe}_S$, respectively. The sketched swimmers have the same size and Brownian diffusivity and only differ in the underlying microscopic swimming dynamics. Variation in the spatial extent of the trajectories is intended to highlight contrasts in the swim diffusivity, while increased thickness of the trajectory-lines reflects larger swim speeds/forces. Schematic (right) of the pivotal role that the swim force plays in materials with attractive interactions. The potential energy displayed is the Morse potential that is utilized in our simulations.}}}
	\label{fig:active-schematic}
\end{figure}

A primary appeal in the use of active dopants is the extraordinary range of swimmers (and properties) that can be exploited. These active particles can be living microscopic organisms such as bacteria or synthetic microswimmers such as catalytic Janus particles. While the mechanism for swimming can be highly variable among swimmers, there are two salient characteristics inherent to all active particles that have proven sufficient in capturing a broad range of properties: (i) the propulsive swim force $\bm{F^{swim}}$ of magnitude $\zeta U_0$ and direction $\bm{q}$, which is the force required to overcome the hydrodynamic resistance to swimming at a speed $U_0$ with a drag coefficient of $\zeta$ and (ii) the characteristic reorientation time $\tau_R$ for an active particle to change its swimming direction $\bm{q}$. The reorientation dynamics of active particles give rise to a persistence length $l_p = U_0\tau_R$ -- the characteristic distance that a particle travels before reorienting -- and, at long times, a swim diffusivity given by $D^{swim} \sim l_p^2/\tau_R = U_0^2\tau_R$.~\cite{Berg1993} 

The strength of advective swimming relative to Brownian diffusion for a particle of size $a$ is given by the swim P\'{e}clet number $\text{Pe}_S = \zeta U_0 a / k_BT$, which can be equivalently thought of as the ratio of the swim force to the characteristic thermal force $\text{Pe}_S = F^{swim}/F^{thermal} = \zeta U_0/(k_BT/a)$. We can also define a  P\'{e}clet number based on the swim diffusivity as $\text{Pe}_R = U_0 a / U_0^2 \tau_R =  a/l_p$.(We use the conventional definition of the P\'{e}clet number as advection over diffusion in defining $\text{Pe}_R$, but others may use the inverse of this quantity.) This reorientation P\'{e}clet number provides a nondimensional measure of the persistence length of the swimmers' motion. Figure~\ref{fig:active-schematic} provides a physical illustration of the influence of $\text{Pe}_S$ and $\text{Pe}_R$ on the microscopic swimming dynamics. We emphasize that active particles with identical long-time properties, such as the swim diffusivity, can have vastly different microscopic dynamics.

Our goal in this work is to use computer simulation to establish generic ``design rules" for the use of active dopants to catalyze material assembly, that is, to establish how the dynamic properties of active dopants ($\text{Pe}_S, \text{Pe}_R$) in relation to the properties of the ``host" material alter the dynamic free energy landscape of the material. To this end, we choose the simplest colloidal system that displays kinetic frustration as our host material: a suspension of polydisperse sticky particles. For this system, the globally stable configuration is macroscopic phase separation (within the two phase boundary) with coexistence between a dilute gas of particles and a dense (glassy) liquid. However, as previously discussed, macroscopic phase separation is kinetically prohibitive, and in practice, a slowly coarsening gel is observed in both experiment and simulation.~\cite{Lu2008, Zaccarelli2008} The clarity in the underlying free energy landscape of this material is precisely what allows us to readily identify the alterations made by active dopants. 

The remainder of the article is organized as follows. First, we use simulation to explore activity space ($\text{Pe}_S, \text{Pe}_R$) and phenomenologically describe the influence of the swimmers. We find remarkably distinct regions in activity space wherein active particles enhance both phase-separation kinetics and mediate effective interactions between the passive particles that can be either repulsive or attractive. This regime in which activity modifies the underlying globally stable configuration is clearly separated from a regime in which active particles only provide a length-scale dependent kinetic enhancement to the coarsening dynamics. We then explore the origins of these regimes and conclude with a discussion on the implication of our results for the use of active dopants for material assembly.

\section{Results and Discussion}

\textbf{Model System.} We consider a system consisting of passive particles of average radius $a$ with a polydispersity of $5\%$ and monodisperse active particles of radius $a$. We primarily focus on two-dimensional systems but selectively perform three dimensional simulations to highlight the generality of our results. The passive particles occupy an area fraction of $\phi_P = 0.4$ while the active area fraction is set to one-tenth of that of the passive particles $\phi_A = 0.04$. All particles undergo translational Brownian diffusion characterized by a self-diffusion time $\tau_B = a^2\zeta/k_BT$. Further, all particles (including the swimmers) interact with the short-ranged attractive potential shown in Figure~\ref{fig:active-schematic} where the well depth is set to $E_0 = 6k_BT$ and the interaction range is $\delta \approx 0.12a$. Note that the active particle interactions need not be the same as those of the passive particles; they are, in principle, entirely independent if the interactions are generated by surface chemistries but will be identical if interactions are due to entropic or geometric effects, such as depletion induced by polymers. Simulations are initialized with random configurations and, unless otherwise noted, are run for a duration of $2000\tau_B$ with the total number of particles ranging from $3300$ to $33000$ (see the Methods for details). 
\begin{figure*}
	\centering
	\includegraphics[width=0.96\textwidth,keepaspectratio,clip]{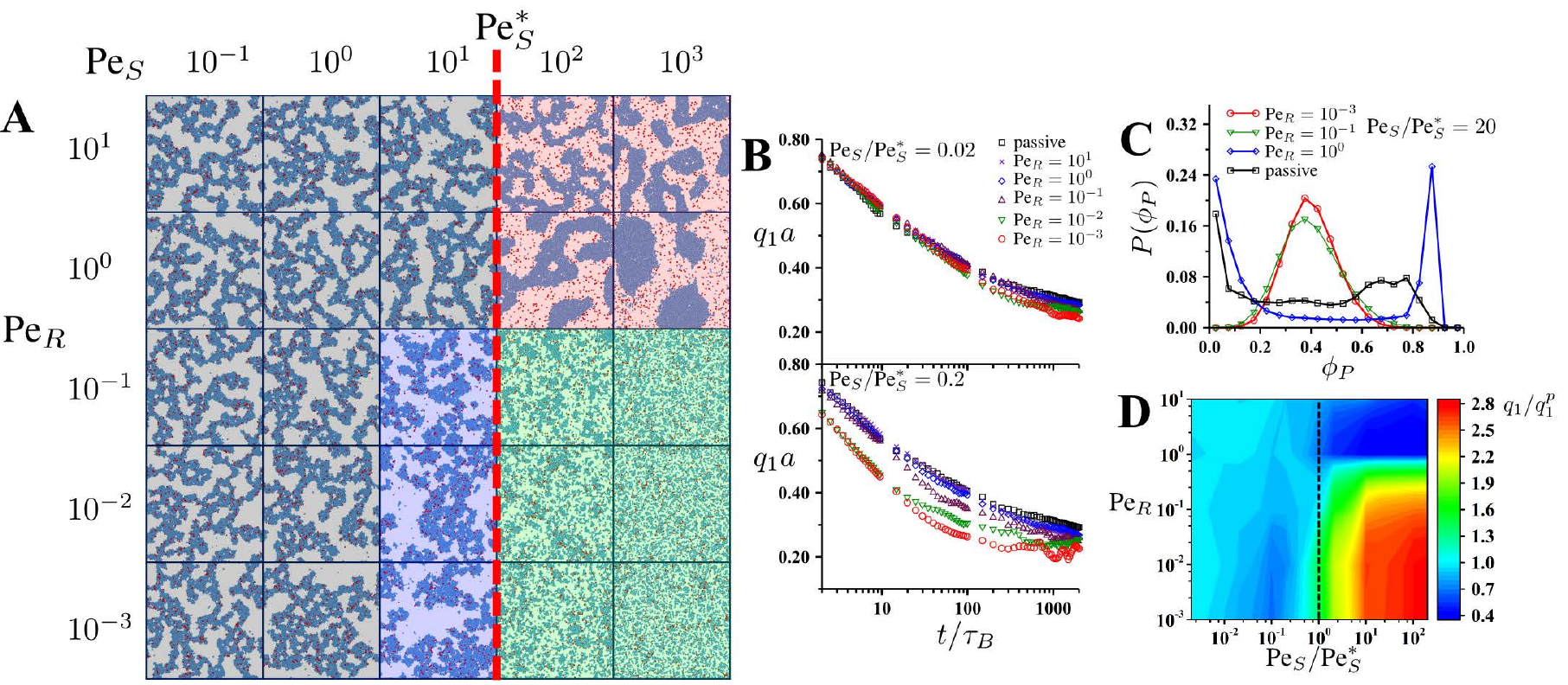}
	\caption{{\protect\small{(A) Representative snapshots of system configurations with like-shaded snapshots representing groupings between dynamically or structurally similar regions in $\text{Pe}_R-\text{Pe}_S$ space. Active particles are shown in red. (B) Dynamic evolution of $q_1$ as a function of persistence as $\text{Pe}_S \rightarrow \text{Pe}_S^*$. (C) Area fraction distribution (obtained by equally dividing the system area into square subareas and computing the local area fraction in each region) as a function of persistence. (D) Dominant (inverse) length scale in the system ($q_1$) as a function of $\text{Pe}_S$ and $\text{Pe}_R$ with interpolation/smoothing applied between $55$ discrete points to achieve a continuous representation.}}}
	\label{fig:phenomenology}
\end{figure*}

As our primary aim is to elucidate the role of activity, we hold all parameters ($E_0$, $\delta$, $\phi_P$, $\phi_A$) fixed while we systematically and independently vary the swimmer persistence length $\text{Pe}_R$ and swim force $\text{Pe}_S$. Before preceding to our simulation results it is import to further contextualize the known role and origins of $\text{Pe}_R$ and $\text{Pe}_S$. Many of the unique active behaviors, such as MIPS~\cite{Fily2012, Cates2015} and particle accumulation on curved surfaces~\cite{Sokolov2010} (to name a few), are entirely characterized by the swimmer persistence length $\text{Pe}_R$.~\cite{Takatori2015, Yan2015} In these cases, the propulsive swim force ($\text{Pe}_S$) can simply be scaled out of the problem. However, when doping a material with active particles we must consider the inherent force scale of the target material. Consider a pair of particles interacting with an attractive potential with a well depth of $E_0$ and a range of a $\delta$. This potential has a natural attractive force scale $F^{a} \sim E_0 /\delta$ that will compete with the swim force of the active particles. When the swim force is smaller than this characteristic attractive force, an active particle will be unable to disrupt the associated particle pair. In contrast, larger swim forces allow swimmers to pull apart interacting particles (see Figure~\ref{fig:active-schematic}). We denote this critical $\text{Pe}_S$ at which the swim force is comparable to the attractive force as $\text{Pe}_S^* = F^{swim*}/F^{thermal}$ with $F^{swim*}=F^a$. The ratio of the swim force to the characteristic attractive force is provided by $\text{Pe}_S/\text{Pe}_S^* = F^{swim}/F^a$, an essential parameter in this study. ($\text{Pe}_S/\text{Pe}_S^* = F^{swim}/F^a$ can be thought of as the analog to the Mason number -- the ratio of shear forces to attractive forces -- in conventional bulk rheology.)

The simple physical consequence described above results in $\text{Pe}_S$ playing a pivotal role in active doping and motivates a careful consideration of the experimentally accessible range of swim forces. Living organisms generate a swim force through cyclic, time-irreversible deformations that induce a straining field in the surrounding fluid and propels the organism.~\cite{Saintillan2018} As the organism uses the fluid to propel, the strength of the swim force is linear in the solvent viscosity so long as the swimmer can continue to generate the same straining field in the fluid and maintain a constant swim speed $U_0$. In practice, the mode of swimming is likely not entirely decoupled from the surrounding fluid viscosity, as living organisms can alter their target swim speeds in response to increased dissipation. However, to a first approximation, $\text{Pe}_S$ can be varied simply by altering the viscosity of the solvent and can therefore be taken as an arbitrarily tunable parameter. 

The flexibility of $\text{Pe}_S$ stands in contrast to the persistence length of biological swimmers which will be dictated primarily by the reorientation mechanism of the organism and in this sense can be viewed as an intrinsic material property of the swimmer. Note that this is not the case for synthetic catalytic Janus particles as their swim speed and, thus, their swim force and persistence can be modulated with the concentration of reactant solute and is reduced with increasing solvent viscosity.~\cite{Anderson1989, BRADY2011} These considerations motivate exploration of a wide range of activity space by \textit{independently} varying $\text{Pe}_R$ and $\text{Pe}_S$ to encompass the extraordinary diverse swimming behavior currently accessible to experiment. This is in contrast to most studies of active matter which assume a thermal reorientation mechanism (true for Janus particles) which couples $\text{Pe}_R$ and $\text{Pe}_S$ (\textit{e.g.},~$\text{Pe}_R = 3/(4\text{Pe}_S)$ in three dimensions) and are thus restricted to a diagonal line in activity space. For a summary of experimentally achievable swim speeds, we refer the reader to the recent review by Bechinger \textit{et al.}~\cite{Bechinger2016} which highlights the extraordinary range of $\text{Pe}_S$ and $\text{Pe}_R$ available in experiments.

\textbf{Structural and Dynamical Control.} Figure~\ref{fig:phenomenology}A illustrates the structural control afforded by active doping. To accentuate the dynamic contrast between regions in activity space, representative snapshots are shown for structures achieved after simulation durations of $2000\tau_B$ and $200\tau_B$ for $\text{Pe}_S \le \text{Pe}_S^*$ and $\text{Pe}_S > \text{Pe}_S^*$, respectively. 

For weak swimming $\text{Pe}_S \ll \text{Pe}_S^* \approx 50$, the active particles are unable to perturb the passive structures -- the initially homogeneous suspension of particles undergoes bicontinuous spinodal decomposition and appears as a coarsening colloidal gel. The coarsening dynamics of the network can be quantified by computing the time evolution of the characteristic largest length scale in the system, which we take as the inverse of the first moment (with respect to the wavevector) of the static structure factor, $q_1(t)$. The weak swimming regime results in coarsening dynamics that are nearly identical to those of a system in which the activity of the swimmers is removed (see Figure~\ref{fig:phenomenology}B). Interestingly, while the active particles in this regime are essentially inert with respect to their kinetic influence, we observe a slight dependence of $q_1(t)$ on the persistence length of the swimmers -- with increasing persistence (decreasing $\text{Pe}_R$) resulting in slightly faster coarsening. This trend holds true but is significantly amplified as we increase the propulsive force toward the critical value $\text{Pe}_S \rightarrow \text{Pe}_S^*$. In this limit, the coarsening dynamics are greatly accelerated with increasing persistence length (Figure~\ref{fig:phenomenology}B) as activity helps drive the coalescence of the gel strands, seen visually in Figure~\ref{fig:phenomenology}A. 

That an active particle is unable to break an individual particle bond yet can assist in the coalescence of network strands (consisting of hundred of bonds) is surprising (and initially somewhat perplexing) and deserves further examination. However, before doing so, we continue with our exploration of $\text{Pe}_S-\text{Pe}_R$ space by now moving to swim forces beyond the critical value $\text{Pe}_{S}^*$. One can visually appreciate that the swimmers are no longer bound to other particles as they can generate the necessary ``escape force" to freely swim. Now, armed with the ability to disrupt the fundamental interaction in the system, the active particles have the capacity to alter the underlying globally stable configuration of the material. 

When the persistence length of the active particles is on the order of their size $\text{Pe}_R \sim O(1)$, the active particles rapidly coarsen the network (the snapshots in Figure~\ref{fig:phenomenology}A represent structures after only $100\tau_B$ of coarsening), approaching complete phase separation on a time scale that is several orders of magnitude faster than a passive system. Moreover, in comparison to the network strands of a passive gel which have an average particle area fraction of $\phi_P \approx 0.8$, the passive regions in this regime are significantly compressed ($\phi_P \approx 0.9$) under the weight of the (osmotic) swim pressure of the exterior active particles as shown in the local area fraction distribution displayed in Figure~\ref{fig:phenomenology}C. This strongly suggests that the swimmers introduce an additional driving force for phase separation, beyond the latent attraction experienced by all particles. (Consistent with this idea, upon simulating particles without any intrinsic attraction (\textit{e.g.},~purely repulsive interactions), we continue to observe (data not shown) phase separation in this regime, which must be entirely driven by the swimmers.)

Upon increasing the persistence we observe a striking inversion of the previous behavior as the active particles now entirely homogenize or ``mix" the system, resulting in the previously bimodal distribution of $\phi_P$ (representative of a phase separated system) becoming normally distributed about the bulk passive area fraction $\phi_P = 0.4$ (Figure~\ref{fig:phenomenology}C). Importantly, if a MIPS-like mechanism were the origin of the phase separation, increasing the persistence $\text{Pe}_R \ll 1$ would deepen the phase separation rather than reverse it.~\cite{Takatori2015, Cates2015, Stenhammar2015, Takatori2015a} Our doping concentration is sufficiently dilute that our observations are truly a reflection of active particles perturbing the host material. 

It is appealing to interpret the $\text{Pe}_S > \text{Pe}_S^*$ regimes as the result of \textit{effective interactions} between the passive particles mediated by the swimmers with the large persistence ``mixing" regime resulting from \textit{swimmer-induced repulsion} and the strongly phase separated regime due to a \textit{swimmer-induced attraction}. While this is highly reminiscent of a depletion-like effect there is a significant caveat: here, our depletant (the active particles) is the \textit{same geometric size} as the passive particles, a fascinating point that we will revisit later.

The boundaries between the four regimes discussed above can be appreciated by plotting $q_1$ of the terminal simulation structure as a function of $\text{Pe}_S$ and $\text{Pe}_R$, shown in Figure~\ref{fig:phenomenology}D. Again, to highlight the dynamic contrast, we plot the $q_1$ achieved after a duration of $2000 \tau_B$ and $200 \tau_B$ (which is often sufficient to reach steady-state for $\text{Pe}_S \gg \text{Pe}_S^*$) for $\text{Pe}_S \le \text{Pe}_S^*$ and $\text{Pe}_S > \text{Pe}_S^*$, respectively. We normalize these $q_1$ values by that of the purely passive system (activity turned off) after a duration of $2000 \tau_B$, which we denote as $q_1^p$. While this plot is the result of $55$ discrete combinations of $\text{Pe}_S$ and $\text{Pe}_R$, the smooth contour assists in the delineation of the four regimes. 

\textbf{Microscale Origins of Dynamic Phase Behavior} The boundary between the high and low persistence behavior for $\text{Pe}_S < \text{Pe}_S^*$ is purely dynamical, while swim forces beyond $\text{Pe}_S^*$ dramatically alter both the dynamics and the globally stable configuration. This clear distinction between regimes that offer kinetic control and those that alter both the kinetics and structure is appealing. The task remains both to understand the physics within these regimes and the nature of the boundaries between them. With the phenomenology established, we now proceed to a systematic exploration on the microscopic mechanisms underpinning each regime of our dynamic phase diagram (summarized in Figure~\ref{fig:dynamicphases}) before concluding with a higher level summary and discussion.

\begin{figure*}
	\centering
	\includegraphics[width=0.65\textwidth,keepaspectratio,clip]{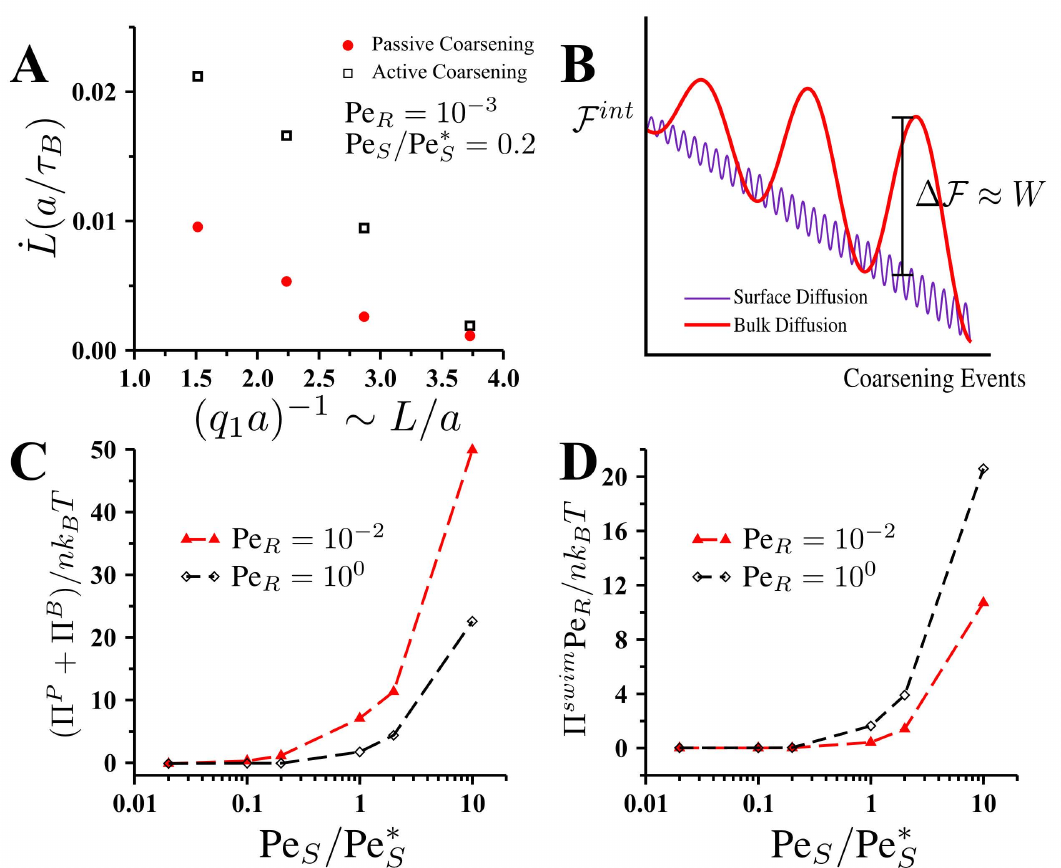}
	\caption{{\protect\small{(A) Coarsening rate for passive and active systems measured as a function of the initial characteristic length scale $L$ of the material and over a duration of $100\tau_B$. The passive coarsening rate was measured by simply ``turning off" the activity of the swimmers in the active system. (B) Schematic of interfacial free energy of a coarsening network as a function of ``coarsening events" for two possible mechanisms. (C) Nonswimming and (D) swimming component of the system pressure as a function of $\text{Pe}_S$ (note that we have scaled out the effect of persistence in panel D). }}}
	\label{fig:microscopic-strain}
\end{figure*}

\textit{Kinetic Control with $\text{Pe}_S < \text{Pe}_S^*$.} We begin by exploring the kinetic pathway provided by active particles with large persistence and $\text{Pe}_S < \text{Pe}_S^*$. The hastening of the gel-coarsening dynamics for these relatively weak activities is consistent with the recent experiments of Szakasits~{\em et al.} which suggests that active particles may enhance the \textit{bulk diffusion} of colloidal gel strands.~\cite{Szakasits2017} In these experiments, a small fraction of active Janus particles were found to increase (by up to a factor of 3) the mean-square displacement (in the subdiffusive regime) of the passive particles comprising the gel despite the activity of the particles being insufficient to break an individual colloidal bond. They rationalized this finding by arguing that the active particles create a long-ranged elastic straining field within the network that displaces the gel strands. That we observe such a significant dependence of the coarsening rate on $\text{Pe}_R$ \textit{suggests} a finite-ranged straining field in the material with a range that correlates with the swimmer persistence length. We postulate that this swimmer-generated strain in the network strands promotes their eventual break-up and coalescence \textit{via} bulk diffusion and is responsible for the observed enhancement of the coarsening dynamics. 

If the swimmers can truly be viewed as microscopic straining agents that promote bulk diffusion, then their influence on the coarsening dynamics should wane as the gel strands thicken. This physical picture is borne out in Figure~\ref{fig:microscopic-strain}A wherein we compare the structural evolution of the active and nonactive system beginning from the same configuration characterized by a length scale of $L \sim q_1^{-1}$. With increasing $L$ the active and passive structural coarsening rates begin to converge. The more rapid decay of the active system coarsening rate (in comparison to the passive system) with increasing strand thickness further (albeit implicitly) corroborates our hypothesis that active particles promote bulk diffusion, which more strongly depends on bulk material properties than surface diffusion.

To further understand activity-assisted coarsening, it is instructive to consider the underlying driving force and mechanism for passive gel coarsening. The gel coarsening is driven by the need to reduce the excess (interfacial) free energy of the system in order to reach the global free energy minimum associated with bulk macroscopic phase separation $\mathcal{F}^{bulk}$. A number of researchers have found that the coarsening of the network strands of passive gels proceeds \textit{via} single-particle diffusion wherein individual particles on the surface of the gel diffuse to energetically more favorable regions of the network.~\cite{Zia2014, Doorn2017} The underlying contributions to the dynamic free-energy barrier in this process consists of both the energetic breaking of the bonds as the particle departs the surface and the entropic barrier associated with the free particle finding an energetically more favorable region. The entropic contribution to this barrier is likely to grow as the gel further coarsens and energetically more favorable regions become more scarce. The surface diffusion mechanism for coarsening can thus be viewed as a series of progressively increasing barriers with each traversed barrier slightly reducing the interfacial area of the material, shown schematically in Figure~\ref{fig:microscopic-strain}B.  

In contrast to the surface diffusion coarsening mechanism, coarsening by bulk diffusion entails the diffusion and coalescence of multiparticle domains, with each of these ``coarsening events" resulting in the reduction of substantially more surface area than in the surface diffusion mechanism. However, the free energy barrier for bulk diffusion is also considerably higher as it involves the breaking of several particle bonds in order for multiparticle regions to break away from the network and diffuse. This primarily energetic barrier will grow as the domains grow thicker (see Figure~\ref{fig:microscopic-strain}) and proves too costly for a traditional passive system to overcome with thermal fluctuations alone, precluding coarsening by bulk diffusion. Passive systems with short-ranged attractive interactions cannot spontaneously generate the stress necessary to overcome the mechanical free energy barrier necessary for bulk diffusion.~\cite{Tanaka2007}

With active dopants, the active particles act as a nonthermal injection of the \textit{mechanical work} necessary to overcome the energetic barriers associated with bulk diffusion in the form stress or pressure -- \textit{e.g.~},``$PV$" work. We can characterize the mechanical influence of the active particles by computing the total pressure $\Pi^T$ of the system
\begin{equation}
\label{eq:pressure}
\Pi^T = \Pi^B + \Pi^P + \Pi^{swim},
\end{equation}
where $\Pi^B$ is the Brownian ``ideal gas" pressure $nk_BT$ (where $n$ is the number density), $\Pi^P$ is the pressure arising from interparticle interactions, and $\Pi^{swim}$ is the swim pressure~\cite{Takatori2014}, the pressure associated with the random walk motion of the active particles. While the swim pressure is small for $\text{Pe}_S < \text{Pe}_S^*$ (as the active particles are precluded from freely swimming), active particles are found to make a substantial indirect contribution to the system pressure reflected in the $\text{Pe}_R$ and $\text{Pe}_S$ dependence of $\Pi^{P}$ (Figure~\ref{fig:microscopic-strain}C). When the swim force is negligible, the overall pressure of the coarsening gel $\Pi^T \approx \Pi^B + \Pi^P$ is very small and, in fact, slightly negative as recently reported.~\cite{Padmanabhan2018} (The physical origin of this negative pressure is that the interfacial area $A^{int}$ and, hence, interfacial free energy of the gel scales with the system size ($\mathcal{F}^{int} \approx \gamma A^{int} \sim V$ where the interfacial tension is positive $\gamma > 0$) and provides a substantial negative contribution ($\Pi^{int} = -\partial \mathcal{F}^{int} /\partial V < 0$) to the total osmotic pressure.) 

As the swim force is increased to its critical value $\text{Pe}_S \rightarrow \text{Pe}_S^*$ the pressure begins to rise sharply, inverting in sign and reaching a value on the order of 10 times $nk_BT$ at $\text{Pe}_S \approx \text{Pe}_S^*$. More persistent particles are found to result in large pressures. Note that when active particles can swim freely the ``ideal" swim pressure is larger for increasing persistence
\begin{equation}
\label{eq:idealswimpressure}
\Pi^{swim}(\phi_A \rightarrow 0) = n_A\zeta U_0^2 \tau_R/2 \sim n_A k_BT\text{Pe}_S/\text{Pe}_R,
\end{equation}
where $n_A$ is the number density of active particles. That the persistence plays such an essential role in generating stress despite the particles being embedded within the material (and unable to freely swim and generate a swim pressure, as shown in Figure~\ref{fig:microscopic-strain}D) further corroborates our assertion that persistence plays a pivotal role in generating the straining field necessary to promote bulk diffusion. It is important to note that despite the large stresses generated when $\text{Pe}_S < \text{Pe}_S^*$, this eventually proves insufficient to overcome the free energy barrier associated with the breakup of increasingly thick gel strands. From this viewpoint, the active particles in this regime can enhance the rate of structure formation up to a particular microscopic length scale. 

\textit{Activity-Mediated Interactions with $\text{Pe}_S > \text{Pe}_S^*$.} We now turn our attention to the $\text{Pe}_S > \text{Pe}_S^*$ region of our dynamic phase diagram. It is interesting to note that for all values of $\text{Pe}_R$ in this region there are no signs of the active particles ``feeling" any attraction. That is, when $\text{Pe}_S > \text{Pe}_S^*$ we observe no systems with statistically significant configurations in which the active particles are clustered in a low-energy state. This physical situation is consistent with the basic intuition offered by many of the phenomenological theories~\cite{Tailleur2008, Wittkowski2014, Takatori2015, Solon2018, Paliwal2018} for active matter ``thermodynamics": active particles have an apparent temperature that is amplified by their intrinsic activity resulting in the leading order ``ideal swimmer" chemical potential (defined by invoking micromechanical or Gibbs-Duhem-like relations using the swim pressure~\cite{Takatori2015, Paliwal2018}) of $\mu^{swim} \sim \zeta U_0^2 \tau_R \ln(\phi_A)$. Thus, despite the vastly different microscopic dynamics of active and passive particles, swimmers can in some cases still be described by an effective temperature given by $\sim \zeta U_0^2 \tau_R$. Caution must be exercised when applying this perspective. For example, while this suggests that active particles with $\zeta U_0^2 \tau_R > E_0$ will behave as hot particles that are unaffected by the weak attraction, swimmers with $\text{Pe}_S < \text{Pe}_S^*$ are unable to generate enough force to escape particle attractions and, as a result, are not spatially distributed in the manner that the effective temperature perspective would suggest. However, as the critical escape swim force is reached, the now \textit{accessible} translational entropy drives the ``release" of active particles from attractive bonds. 

The freely swimming particles with $\text{Pe}_S > \text{Pe}_S^*$ entirely reshape the globally stable configuration of the material, pushing the system to disparate regions of phase space: complete homogenization at larger persistence lengths ($\text{Pe}_R \ll 1$) or strong phase separation with persistence lengths on the order of the body size of the swimmer. These behaviors motivate the perspective that the active particles are mediating effective (\textit{e.g.},~depletion) interactions between the passive particles. Numerous studies have sought to describe activity-mediated interactions between passive particles~\cite{Angelani2011, Ray2014, Parra-Rojas2014, Harder2014, Ni2015, Smallenburg2015} with most of these studies fixing two passive objects in an active bath and measuring the force required to keep the objects fixed, the original perspective taken by Asakura and Oosawa (AO) in describing the origins of traditional passive depletion.~\cite{Asakura1954, Asakura1958} However, the role of the size-ratio of the depletant and passive body -- which is decisive in traditional thermal depletion~\cite{Lekk2011} -- remains largely unexplored. The original AO picture is that \textit{smaller} particles (the depletants) will create an entropic attractive force between \textit{larger} particles in an effort to increase the space available to the smaller particles and lower the system free energy. However, our active particles are of the \textit{same geometric size} as our passive particles. In the context of the previously discussed ``effective" translational entropy, the swimmer-induced attraction at larger $\text{Pe}_R$ can be thought of as depletion attraction driven by an \textit{activity asymmetry} with no geometric size difference required.

\begin{figure}
	\centering
	\includegraphics[width=0.48\textwidth,keepaspectratio,clip]{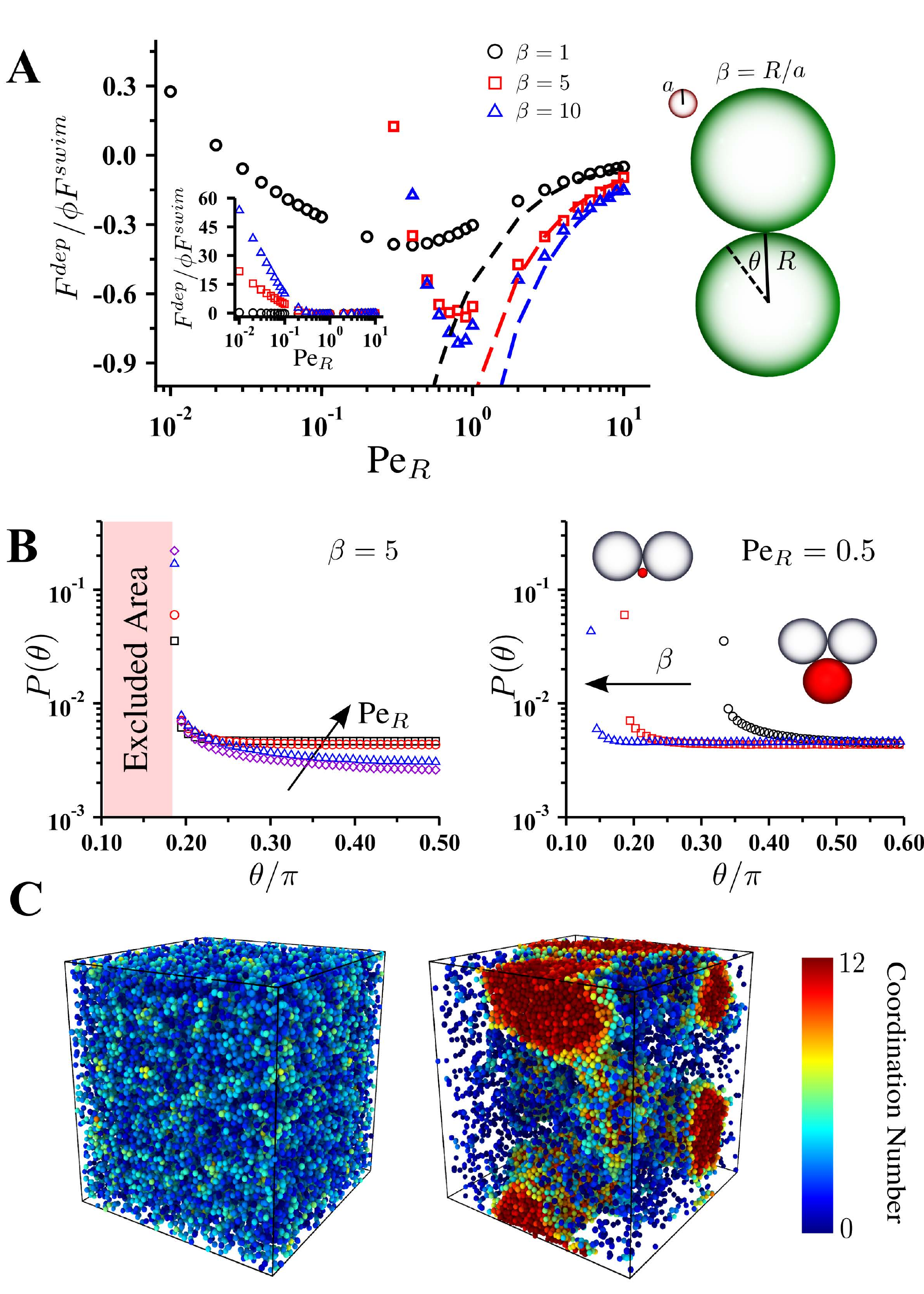}
	\caption{{\protect\small{(A) The probe-swimmer size-ratio and swimmer persistence length dependence of the mechanical depletion force with the lines representing eq~\eqref{eq:AOforcedimen}. The nonmonotonic region is magnified here, while the inset displays the complete data in its entirety. (B) Swimmer-probe contact statistics as a function of the (left) persistence (with $\text{Pe}_R = 1, 0.5, 0.1, 0.05$) and (right) size-ratio  (symbols are the same as in panel A). (C) Representative snapshots after $100\tau_B$ of a three-dimensional system of monodisperse active particles (not shown) and polydisperse ($5\%$) passive particles at volume fractions of $\phi_A = 0.02$ and $\phi_P  =0.20$, respectively, with $\text{Pe}_S/\text{Pe}_S^* = 200$ and (left) $\text{Pe}_R = 10^{-2}$ and (right) $\text{Pe}_R = 1$. Particles are colored by coordination number to aid in cluster visualization.}}}
	\label{fig:depletion}
\end{figure}

The concept of an activity asymmetry driven depletion interaction merits further systematic examination and will invariably require relaxing many of the underlying assumptions in the traditional AO picture. For example, when the ratio of the  depletant size (radius $a$)  to the passive body size (radius $R$) (see schematic in Figure~\ref{fig:depletion}A) exceeds a value of $a/R > 0.15$, the depletion interaction is known to no longer be pairwise, with the resulting multibody interactions deepening the effective attraction between the passive bodies.~\cite{Lekk2011} Furthermore, the classical AO picture of treating the depletant as ideal will invariably break down with larger depletant particles, which are subjected to excluded-volume-related correlations. Nevertheless, in an effort to isolate the leading order role of the size-ratio in active depletion, here we explore only the two-body interaction between passive particles in an ideal active bath. We fix two identical passive disks at contact in a bath of non-Brownian (to isolate the role of activity) active particles at an area fraction of $\phi$. In the limit of infinitesimal persistence length ($\text{Pe}_R \rightarrow \infty$), the active particles behave as Brownian particles with the force acting between the passive bodies \textit{at contact} given by an AO form
\begin{equation}
\label{eq:AOforce}
F^{dep} = -\frac{\zeta U_0^2 \tau_R \phi}{a\pi}\sqrt{1+2\beta},
\end{equation}
where $\beta = R/a$ is the size ratio, and we have substituted the activity $\zeta U_0^2 \tau_R/2$ in for what is traditionally $k_BT$ (an exact substitution in the limit $\text{Pe}_R \rightarrow \infty$). (Note that this idealized AO perspective predicts a finite attractive force for \textit{all} size ratios. This is clearly incorrect and is the result of neglecting the concentration and translational entropy loss of the passive bodies.) The active-depletion force must necessarily scale as $F^{dep} \sim \phi F^{swim}$ as the active bath is dilute and the swim force is the only force-scale in the problem. The nondimensional depletion force $F^{dep}/\phi F^{swim}$ is thus entirely determined by geometric considerations (size ratio $\beta$ and persistence length $\text{Pe}_R$) with a value of
\begin{equation}
\label{eq:AOforcedimen}
F^{dep}/\phi F^{swim} = -\frac{\sqrt{1+2\beta}}{\pi\text{Pe}_R},
\end{equation}
to leading order in $\text{Pe}_R^{-1}$. Equation~\eqref{eq:AOforcedimen} has a number of noteworthy features: (i) the depletion force is always attractive (negative); (ii) the magnitude of the force is a monotonically increasing (decreasing) function of persistence ($\text{Pe}_R$); and (iii) the magnitude is an increasing function of the size ratio $\beta$.

Independent of size ratio, as the persistence increases we find a nonmonotonic response in the depletion force (shown in Figure~\ref{fig:depletion}A) with eq~\eqref{eq:AOforcedimen} serving as a lower bound as was found by Smallenburg and L{\"o}wen.~\cite{Smallenburg2015} Initially, decreasing $\text{Pe}_R$ results in an increased attraction (more negative) between the particles, consistent with the AO expectations: the particles are ``hotter" and should result in the force decreasing with $\text{Pe}_R^{-1}$ (eq~\eqref{eq:AOforcedimen}). However, with increasing persistence, the particles begin to accumulate at the wedge created by the two passive disks, with the distribution of swimmers on the passive disk surfaces transitioning from homogeneous (with the exception of the exclusion zone) to sharply peaked at the wedge (see Figure~\ref{fig:depletion}B). This results in a repulsive force that competes with the traditional AO attractive force, resulting in the depletion force reaching a minimum followed by a sharp increase -- particularly for $\beta > 1$. Interestingly, while the traditional AO picture (eq~\eqref{eq:AOforcedimen}) suggests that larger depletants reduce attraction -- and indeed, we find the minimum force is an increasing function of size ratio -- here we find that reducing the size ratio $\beta$ significantly broadens the range of $\text{Pe}_R$ where the depletion force is attractive, with an order of magnitude increase in the persistence length at which the attraction inverts to repulsion. This stems from the reduction in swimmer accumulation within the wedge with increasing swimmer size -- larger swimmers are less likely to be ``trapped" in the wedge as illustrated in Figure~\ref{fig:depletion}C. Thus, like-size swimmer depletants not only reveal and emphasize the important notion of activity asymmetry driven interactions but fundamentally broaden the window of persistence lengths in which swimmer-mediated attraction is observed. 

The effective repulsion found when the persistence length is large compared to the passive body size ($R/l_p = \text{Pe}_R\beta \ll 1$) is in agreement with the homogenization observed in our full mixture simulations at small $\text{Pe}_R$. This picture -- while mechanically intuitive -- results in the active particles conceding free volume, a result which appears at odds with the intuition offered by phenomenological active-matter thermodynamic theories previously referenced.~\cite{Tailleur2008, Wittkowski2014, Takatori2014} While interesting, reconciling this is beyond the scope of this work.

We emphasize that the depletion physics described above should hold for three dimensions as well. We explicitly verify this by simulating the 3D analogue of our system in the regimes of swimmer-induced repulsion ($\text{Pe}_S > \text{Pe}_S^*$ and $\text{Pe}_R \ll 1$) and attraction ($\text{Pe}_S > \text{Pe}_S^*$ and $\text{Pe}_R \sim O(1)$), with representative snapshots of terminal configurations shown in Figure~\ref{fig:depletion}C (see the figure caption for simulation details). We indeed find activity-induced rapid homogenization/phase separation in the expected regimes. 

\begin{figure}
	\centering
	\includegraphics[width=0.48\textwidth,keepaspectratio,clip]{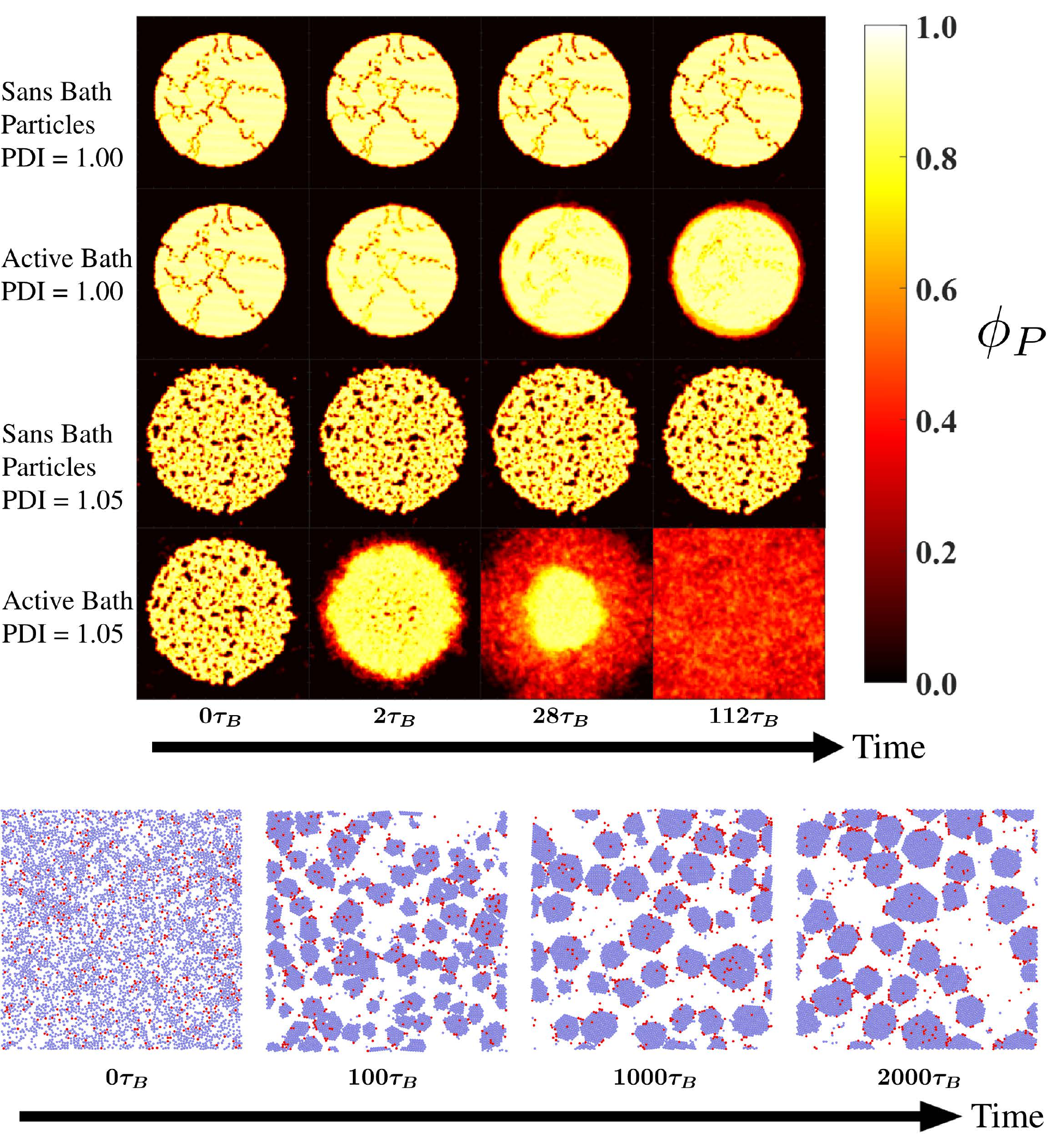}
	\caption{{\protect\small{ (Top) Rows represent the time-evolution of a prepared passive-disk consisting of either monodisperse or polydisperse particles with and without surrounding active particles with $\text{Pe}_S/\text{Pe}_S^* = 1.25$ and $\text{Pe}_R=10^{-2}$ (\textit{e.g.},~the activity-induced repulsion regime). (Bottom) Time evolution of a monodisperse system with identical parameters as the above disk simulations but beginning from a spatially homogeneous configuration. Active particles are shown in red. }}}
	\label{fig:microstructure}
\end{figure}

\textit{Microstructural Dependence of Active Depletion.} We argued above that the boundary delineating the activity-induced attraction and repulsion regimes is entirely dictated by $\text{Pe}_R$ and $\beta$ -- with $\beta$ being the size ratio between the passive and active particles. That we can think of the active particles as interacting with individual passive particles suggests that the swimmers have sufficient propulsive force to ``dislodge" the passive particle from whatever latent structure the particles form. This is indeed the case for our system when the propulsive swim force exceeds the critical value $\text{Pe}_S^*$. However, this may not generally hold as, in principle, the latent microstructure of the passive particles may prove too formidable for an active particle to disrupt. To illustrate this point, we prepare two systems that nominally differ only in the local microstructure by varying the passive particle size distribution. We initialize our systems by arranging all of the passive particles into a close-packed disk at the center of the simulation cell, with one system consisting of polydisperse ($5\%$) passive particles and the other consisting of monodisperse passive particles. Both disks are stable over time in the absence of active particles as macroscopic phase separation is the globally stable configuration. The monodisperse system is crystalline (with some defects), while the polydisperse system is disordered with a number of large defects (see Figure~\ref{fig:microstructure}). We next insert active particles into the open space outside of the disks. With $\text{Pe}_S = 1.25\text{Pe}_S^*$ and $\beta \text{Pe}_R \ll 1$, we anticipate that the active particles should induce an effective repulsion between particles and homogenize the system, which is indeed true for the polydisperse system. However, for the crystalline system no such mixing was observed. Rather, we find that active particles build up on the surface -- they are unable to penetrate the material -- and compress the disk (the details regarding the transience of this compression were explored by Stenhammar \textit{et al.}~\cite{Stenhammar2015}) as shown in Figure~\ref{fig:microstructure}. 

That the active particles appear unable to induce repulsion between the monodisperse particles stems from their inability to dislodge individual particles from the crystalline microstructure. However, this does not mean that the swimmers are unable to induce effective interactions. Rather, we argue that the crystallinity alters the interaction from the expected repulsion to attraction. This implies that the structures that active particles mediate interactions between are not necessarily individual particles, but rather the smallest length scale structure that the swimmers can ``push" -- an implicitly necessary criteria for depletion-like interactions. The ratio of the length scale of this intrinsic structural unit $L_0$ to the swimmer size $a$ is the relevant size ratio (along with $\text{Pe}_R$) in determining the strength/sign of the active depletion force. 

This length scale $L_0$ is sensitive to the underlying microstructure (characterizable by the passive-particle pair distribution function $g_0(r)$) that the swimmer must compete with, the swim force $\text{Pe}_S$, and perhaps even the preparation conditions of the material. Consider the monodisperse system but now prepared from a homogeneous initial configuration. Nearly instantly after we begin the simulation, the passive particles nucleate into several small crystalline ``islands" that the swimmers cannot penetrate (see Figure~\ref{fig:microstructure}). Initially when the island size $L_0$ is small compared to the persistence length, the swimmers induce repulsion between islands as $L_0/l_p = \beta \text{Pe}_R \ll 1$. However, over time islands eventually collide and merge, with the swimmers unable to break apart the resulting structure. The average island size $L_0$ will \textit{eventually} become much larger than the persistence length ($L_0/l_p \gg 1$) as activity and thermal energy are unable to break the islands. In this limit, the swimmers would induce interisland attraction and \textit{facilitate} island coalescence. Intriguingly, the influence of ``microstructural integrity" (which the polydisperse system lacks) is to \textit{expand} the window of activity-induced attraction by driving up the effective size ratio $\beta_{eff} = L_0/a$ where $L_0$ is a functional of $g_0(r)$, $\text{Pe}_S$, and possibly the preparation history of the material. 

We caution that \textit{only} when the persistence length is much smaller than all of the other geometric length scales of the passive bodies ($L_0/l_p \gg 1$) is activity-mediated attraction \textit{ensured}. It is precisely in this limit wherein the active particles become equivalent to ``hot" Brownian particles and traditional AO physics is applicable (\textit{e.g.},~eq~\eqref{eq:AOforcedimen}). Short of this limit, geometric details -- which may strongly influence the distribution of swimmers on the passive body surface -- must be considered, as highlighted in Figure~\ref{fig:depletion}A.
  
\begin{figure}
	\centering
	\includegraphics[width=0.48\textwidth,keepaspectratio,clip]{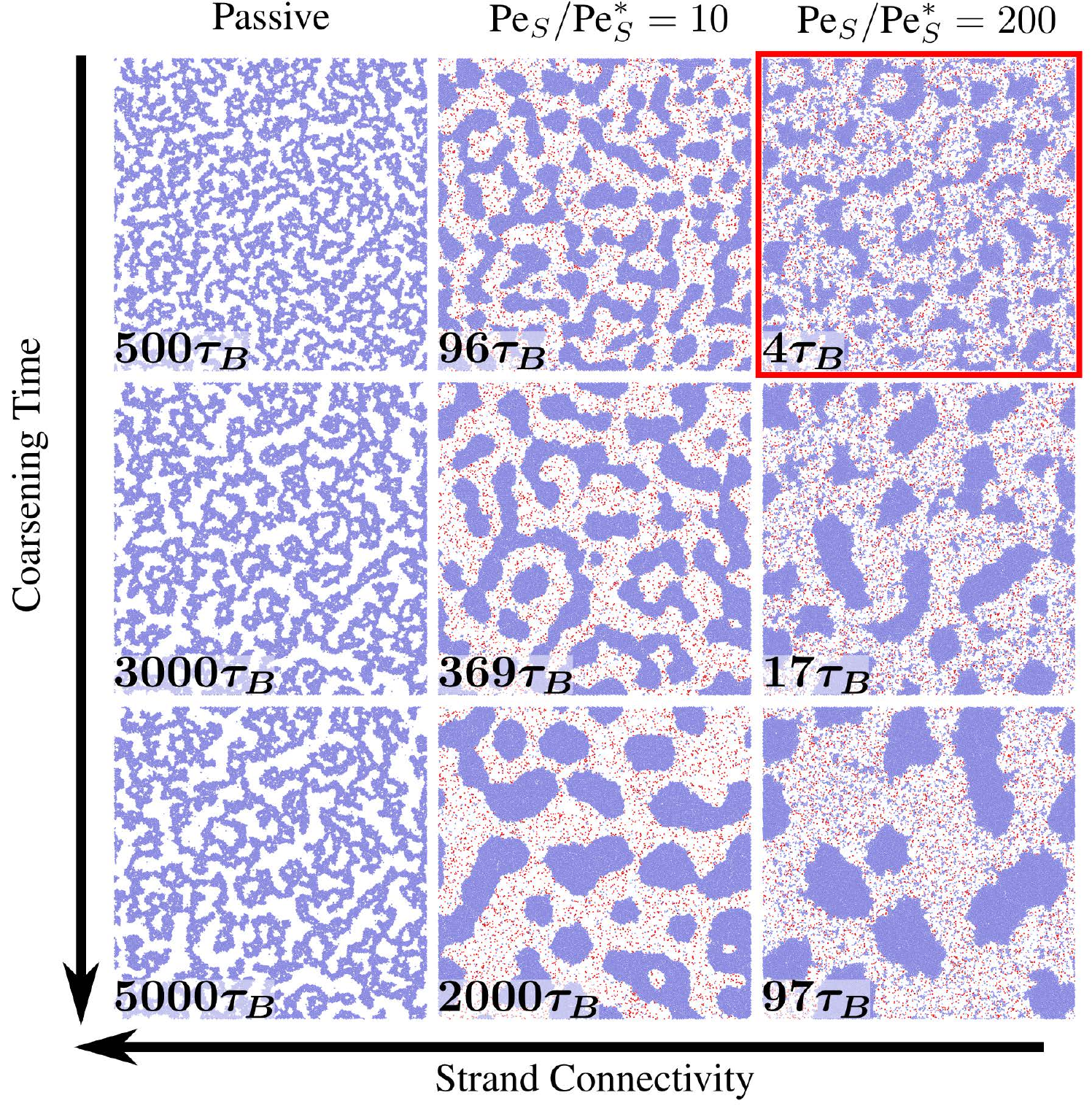}
	\caption{{\protect\small{Qualitative illustration of both the rapid coarsening in the activity-induced attraction regime and the transition from bicontinuous to droplet-like spinodal decomposition with increasing $\text{Pe}_S$ for $\text{Pe}_S > \text{Pe}_S^*$ and $\text{Pe}_R = 1$. Active particles are shown in red. The left-most column illustrates the coarsening process for a passive system.}}}
	\label{fig:coarsening}
\end{figure}

\textit{Dynamic Consequences of $\text{Pe}_S > \text{Pe}_S^*$.} With $\text{Pe}_S > \text{Pe}_S^*$, the active particles not only modify the effective interactions between the passive particles, but also dramatically alter the kinetics of phase separation. Figure~\ref{fig:coarsening} illustrates representative snapshots of the coarsening process in the activity-induced attraction regime. Note that here the system size (number of particles) is 10 times larger than the systems shown in Figure~\ref{fig:phenomenology}A and the growth of the domains is monotonic in time, indicative of true macrophase separation. While increasing the magnitude of attraction is typically thought to slow particle dynamics, with increasing $\text{Pe}_S/\text{Pe}_S^*$ we find an increase in the coarsening rate of the domains. This highlights the important point that, dynamically, active particles can ``share" their diffusivity with the passive particles: providing both strong attractive interactions and the necessary ``kicks" to rapidly assemble the material. The material is structurally cool but dynamically hot. 

Fascinatingly, as we increase the activity-induced attraction by increasing $\text{Pe}_S$ further beyond $\text{Pe}_S^*$, the connectivity of the strands begins to decrease, with the dense phase no longer a bicontinuous percolating cluster (a gel) at the largest $\text{Pe}_S$ explored. While strand connectivity is of course expected to be reduced with increasing strand thickness, we emphasize that this change in the coarsening process is observed even in the early stages of phase separation (see the snapshot enclosed in a red square in Figure~\ref{fig:coarsening}). This suggests a fundamental alteration of the coarsening mechanism from bicontinuous spinodal decomposition -- wherein a gel-like space-spanning structure thickens with time -- to a mechanism that appears similar to droplet spinodal decomposition (see Figure~\ref{fig:coarsening}), where demixing is achieved by the rapid formation of particle-rich drops that merge over time.~\cite{Tanaka2000} 

For traditional passive systems, transitioning from bicontinuous to droplet spinodal decomposition is expected to only occur by modulating the global particle concentration; the different behavior observed here highlights the crucial dynamic role of activity. It was recently shown that the mechanism for droplet coalescence is due to Marangoni-type forces that induce an interdroplet correlation in addition to typical Brownian diffusion.~\cite{Shimizu2015} The rapid droplet coalescence observed in our simulations is likely due to the nonthermal diffusion provided by the swimmers to the droplets, in a manner similar to the oft-studied tracer diffusion problems in active matter.~\cite{Peng2016, Burkholder2017} It will be interesting to see if the scalar active matter theories~\cite{Wittkowski2014} used to explore the phase separation dynamics of purely active systems can capture the coarsening behavior reported here.

\section{Conclusions}

\begin{figure}
	\centering
	\includegraphics[width=0.48\textwidth,keepaspectratio,clip]{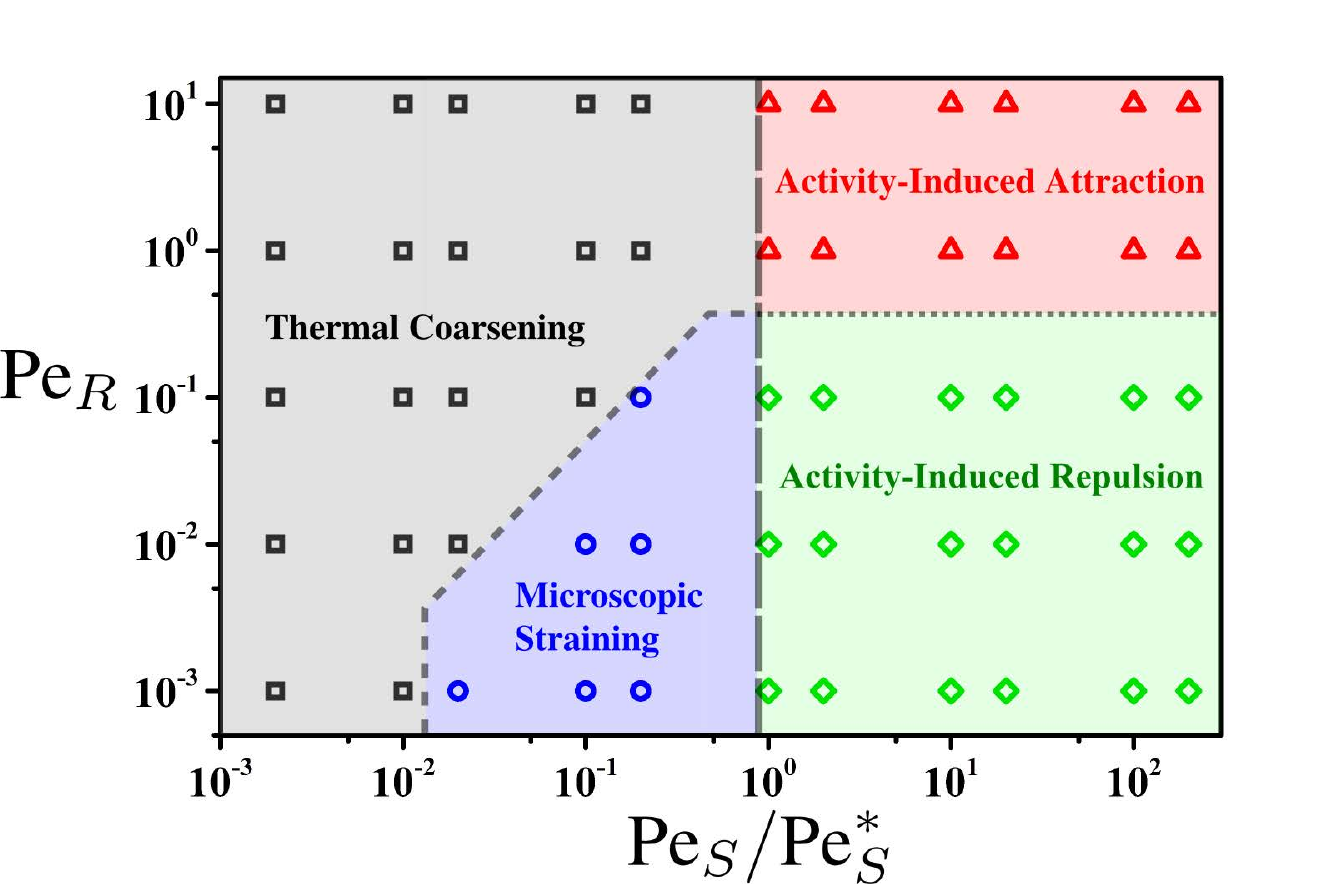}
	\caption{{\protect\small{Dynamic phase diagram of an active doped gel. Materials embedded with active particles in the microscopic straining regime will exhibit improved coarsening rates on length scales $L$ that correlate with swimmer persistence $l_p$. The regimes of activity-induced attraction/repulsion are due to depletion-like forces and are separated by a geometrically controlled boundary that depends on $\beta_{eff}$ and $\text{Pe}_R$. }}}
	\label{fig:dynamicphases}
\end{figure}

We explored the use of active dopants for material design using a simple model system of a suspension of gel-forming sticky particles seeded with a small fraction of swimmers. We found that the active dopants provide a wide range of structural and dynamical control as a function of the swim force ($\text{Pe}_S$) and persistence length ($\text{Pe}_R$), the two properties governing the underlying microscopic swimmer dynamics. By systematically exploring the full $\text{Pe}_S-\text{Pe}_R$ space, we observed three distinct regimes afforded by the active dopants: microscopic straining, activity-induced attraction, and activity-induced repulsion, as summarized in the dynamic phase diagram presented in Figure~\ref{fig:dynamicphases}. We stress that the boundaries have a degree of ``fuzziness" as there are finite transition zones, perhaps best represented in Figure~\ref{fig:phenomenology}D. Nevertheless, the distinctions between the regimes are clear. 

In the microscopic straining regime, the swim force is near its critical value ($\text{Pe}_S \rightarrow \text{Pe}_S^*$), allowing swimmers to strain the material structure and provide kinetic control at length scales $L$ comparable to the persistence length $l_p$ of the particles, a feature that is perhaps attractive for microphase separating systems. Here, the swimmers are unable to generate enough force to fundamentally alter the underlying globally stable configuration of the material. In contrast, when the propulsive swim fore is increased beyond its critical value $\text{Pe}_S^*$, the swimmers mediate strong effective interactions that entirely alter the material phase behavior while also fundamentally altering the system dynamics. 

In this regime, active particles are found to act as depletants despite there being no geometric size asymmetry between them and the passive particles, revealing the notion of ``activity asymmetry" driven depletion. The effective attraction regime is broadened with increasing ``microstructural integrity" with coarsening dynamics that are fundamentally altered in comparison to analogous passive systems. While the effective attraction mediated by the swimmers will deepen the phase separation, the remarkable kinetics afforded by this regime introduces the intriguing prospect of using activity to quickly complete the phase separation process with a resulting configuration that is ``supercooled" with respect to the globally stable configuration in the absence of activity. This idea would be particularly applicable to materials in which the underlying globally stable configuration is not fundamentally altered with increasing attraction, as is the case with many close-packed structures. Activity could then simply be ``switched off" (\textit{e.g.},~by ceasing to provide the necessary fuel for the active particles to swim) with the resulting structure achieved on a time scale that is orders of magnitude smaller than passive dynamics would allow for.  

The clear delineation between regimes in which active particles preserve the underlying globally stable configuration ($\text{Pe}_S < \text{Pe}_S^*$) and those in which swimmers entirely alter the underlying phase behavior ($\text{Pe}_S > \text{Pe}_S^*$) is an enticing feature of active dopants. That the swim force plays the decisive role in dictating the location on the phase diagram results in a number of intriguing effects. Principal among these is that simply by increasing the system viscosity (in a manner that does not alter the particle chemistry) one can modulate the swim force exerted by many biological microswimmers and can therefore dramatically alter the phase behavior of the entire material, \textit{e.g.}, a macroscopic transport property can alter the state or phase of a system.

Active doping for material design has a number of appealing features. In contrast to macroscopic deformation protocols, active doping does not intrinsically break the rotational symmetry of the system and can therefore be used to improve the self-assembly of ordered structures with various symmetries. Further, the properties of the swimmers are in principle entirely orthogonal to those of the host material. It is our hope that by establishing the dynamic phase diagram for a model material and elucidating the key length, force and time scales delineating the dynamic phase boundaries, our results can be directly tested in experiment and can be applied quite generally to a host of colloidal materials with varying chemistries, interaction potentials, and free energy landscapes. While we have focused primarily on the implications of our work for material design, it would be intriguing to probe the applicability of our results toward naturally occurring systems of swimming microorganisms embedded in ``sticky" gel-like environments such as biofilms.~\cite{Flemming2010} Our findings suggest that, armed with a sufficient amount of activity and depending on the length/force scales of the surroundings, microorganisms can fundamentally reshape their environment -- an interesting conjecture that merits further examination. 

\section{Methods}
\footnotesize
We model our passive particles as polydisperse disks with a mean radius $a$ and a polydispersity of $5\%$. The polydispersity is included by normally distributing the particle radii between $21$ discrete sizes that are evenly spaced apart with a size range of $a \pm 15\%a$. The drag coefficient $\zeta_i$ of particle $i$ is linearly scaled with the particle size $a_i$. The active particles are modeled as monodisperse disks with radius $a$. Initial configurations are generated by placing the particles randomly followed by applying the potential-free algorithm~\cite{Heyes1993} to move overlapping disks to contact. 

The motion of the particles is governed by the overdamped Langevin equation which, upon integration over a discrete time step $\Delta t$, results in the displacement equation
\begin{equation}
\label{eq:displacementeq}
\Delta \bm{x_i} =\Delta \bm{x^B_i} + \frac{1}{\zeta_i}\left(\bm{F^{swim}_i} + \sum_{j\ne i}^{}\bm{F^P_{ij}} \right)\Delta t, 
\end{equation}
where $\Delta \bm{x_i}$ is the particle displacement over the simulation time step and $\Delta \bm{x^B_i}$ is the stochastic Brownian displacement (taken to be a white noise with a mean of $\bm{0}$ and variance of $2D^B\Delta t\bm{I}$ with $D^B = k_BT/\zeta_i$), while $\bm{F^{swim}_i}$ and $\bm{F^P_{ij}}$ are the swim force and interparticle forces from particle $j$, respectively. For passive particles, the swim force is identically zero, while for active particles it is defined as $\bm{F^{swim}_i} = \zeta U_0\bm{q_i}$. We model the reorientation dynamics of $\bm{q_i}$ as a continuous random process (as opposed to, for example, run-and-tumble motion) with the magnitude of random torques selected such that the rotational diffusivity is $\tau_R^{-1}$. In two dimensions, $\bm{q_i} = (\cos(\theta_i), \sin(\theta_i))$ where the angle $\theta_i$ evolves according to
\begin{equation}
\label{eq:displacementeorient}
\Delta \theta_i = \frac{1}{\zeta^R_i} L^R_i\Delta t, 
\end{equation}
where $\Delta \theta_i$ is the angular displacement over the simulation time step, $\zeta^R_i$ is the rotational drag, and $L^R_i$ is a random torque with a mean of $0$ and a variance of $2 (\zeta^R_i)^2/(\tau_R\Delta t)$. Note that the rotational drag has no dynamical consequence as eq~\eqref{eq:displacementeorient} can be rewritten as $\Delta \theta_i = \tilde{L}^R_i\Delta t$ where the redefined torque $\tilde{L}^R_i$ now has a variance of $2/\tau_R\Delta t$. Note that we do not assume that the rotational diffusion is thermal in origin, and thus, $\tau_R$ is entirely decoupled from all other system parameters. The interparticle force $\bm{F^P_{ij}} = - \partial E_{ij}(r_{ij}) / \partial \bm{r_{ij}}$ derives from the Morse potential
\begin{equation}
\label{eq:potential}
E_{ij}(r)/E_0 = \exp[-2\kappa (r-d_{ij})] - 2\exp[-\kappa (r-d_{ij})],  
\end{equation}
where $d_{ij}$ is the sum of the radii of the interacting particles $d_{ij} = a_i + a_j$ and $\kappa$ is an inverse screening length set to $\kappa a = 30$. Pair interactions are cutoff beyond distances of $1.25d_{ij}$. The simulation time step must be reduced with increasing particle activity to ensure numerical stability of eq~\eqref{eq:displacementeq}. For $\text{Pe}_S/\text{Pe}_S^* \le 2$, we use a time step of $\Delta t = 5\times10^{-6}$, while for $\text{Pe}_S/\text{Pe}_S^* = 10$, $20$, $100$, and $200$ we use time steps of $\Delta t = 1\times 10^{-6}$, $5 \times 10^{-7}$, $1\times 10^{-7}$, and $5\times 10^{-8}$, respectively.

The large-scale simulations conducted for the data in Figure~\ref{fig:microscopic-strain}C, D, Figure~\ref{fig:depletion}C, and Figure~\ref{fig:coarsening} were performed using the GPU-enabled HOOMD-blue molecular dynamics package with a total of $33000$ particles.~\cite{Anderson2008, Glaser2015} All other data were obtained using our in-house CPU-based code with the simulations presented in Figure~\ref{fig:phenomenology} and Figure~\ref{fig:microscopic-strain}A consisting of $3300$ particles and the simulations presented in Figure~\ref{fig:microstructure} consisting of $5500$ total particles.

We obtained the simulation results in Figure~\ref{fig:depletion} by using the potential-free algorithm to model truly hard spheres.~\cite{Heyes1993} The force required to hold a probe particle fixed is simply the sum of the collisional forces~\cite{Foss2000} imparted by the active bath particles. For these simulations, we select a time step of $\Delta t = 10^{-3}a/U_0$. Simulations were run for a duration of $1000\tau_R$ with $10000$ independent swimmers. Swimmers can induce incredibly long-ranged ($\sim O(l_p)$) interactions between passive probes, requiring the simulation box size to be made sufficiently large to prevent self-interactions of the fixed probes with their periodic images.

The characteristic largest length scale in the simulation was taken to be inversely proportional to the first moment of the static structure factor $q_1$
\begin{equation}
\label{eq:q1}
q_1(t) = \frac{\int_{q_{m}}^{q_{c}}S(q;t)qdq}{\int_{q_{m}}^{q_{c}}S(q;t)dq},  
\end{equation}
where $q_{m}$ is the minimum possible wavenumber (physically bounded by the size of our simulation cell) and $q_{c}$ is the cutoff wavenumber which we set to $2q_{c}a = 3$ to include the contribution of all large wavelengths as was done in the experiments of Lu \textit{et al}.~\cite{Lu2008} The static structure is computed directly from the particle coordinates
\begin{equation}
\label{eq:S(q)}
S(q;t) = \frac{1}{N}\left \langle \left \lvert \sum_{j=1}^{N}\exp(i\bm{q}\cdot{\bm{x_j}(t)}) \right \rvert^2 \right \rangle,  
\end{equation}
where $N$ is the total number of \textit{passive} particles.

The interaction component to the pressure $\Pi^P$ was computed using the standard virial approach with $\Pi^P = n \langle \bm{x_{ij}}\cdot \bm{F_{ij}} \rangle/2$ where $\bm{x_{ij}}$ and $\bm{F_{ij}}$ are the interparticle distance and force between particles $i$ and $j$, respectively, $n$ is the number density of the system, and the brackets denote an ensemble average over all particle pairs. We compute the swim pressure using the ``impulse" expression~\cite{Solon2015a} that is valid for active particles which feel no interparticle or external torques
\begin{equation}
\label{eq:swimimpulse}
\Pi^{swim} = n\left \langle \tau_R\bm{v_i} \cdot \bm{F^{swim}_i}\right \rangle/2,  
\end{equation}
where $\bm{v_i}$ is the net velocity of particle $i$ (\textit{e.g.},~eq~\eqref{eq:displacementeq} divided by $\Delta t$). This expression results exactly in the correct swim pressure for ideal active particles and has been shown~\cite{Patch2018} to reproduce the traditional virial method~\cite{Takatori2014, Fily2014} for computing the swim pressure.
\section{Acknowledgments}
A.K.O. acknowledges support by the National Science Foundation Graduate Research Fellowship under Grant No. DGE-1144469 and an HHMI Gilliam Fellowship. Y.W. acknowledges the Caltech SURF program and support by the Talent Training Program in Basic Science of the Ministry of Education of the People's Republic of China. J.F.B. acknowledges support by the National Science Foundation under Grant No. CBET-1437570.

\providecommand{\latin}[1]{#1}
\makeatletter
\providecommand{\doi}
{\begingroup\let\do\@makeother\dospecials
	\catcode`\{=1 \catcode`\}=2 \doi@aux}
\providecommand{\doi@aux}[1]{\endgroup\texttt{#1}}
\makeatother
\providecommand*\mcitethebibliography{\thebibliography}
\csname @ifundefined\endcsname{endmcitethebibliography}
{\let\endmcitethebibliography\endthebibliography}{}

\end{document}